# Impact of Casimir Force in Molecular Electronic Switching Junctions

Aaron Katzenmeyer[1], Logeeswaran VJ[1], Bayram Tekin[2] & M. Saif Islam[1]

[1]Department of Electrical & Computer Engineering
University of California at Davis, Davis CA 95616 USA
Email: saif@ece.ucdavis.edu

[2]Department of Physics
Middle East Technical University, 06531 Ankara, Turkey

*Abstract-* **Despite significant progress in synthesizing several new molecules and many promising single device demonstrations, wide range acceptance of molecular electronics as an alternative to CMOS technology has been stalled not only by controversial theories of a molecular device's operation, for example the switching mechanism, but also by our inability to reproducibly fabricate large arrays of devices. In this paper, we investigate the role of Casimir force as one of the potential source of a wide range of discrepancies in the reported electrical characteristics and high rate of device shorting in molecular electronic switching junctions fabricated by sandwiching a molecular monolayer between a pair of planar metal electrodes.**

I. INTRODUCTION

Although the Casimir force is known throughout the theoretical physics community, it has only recently been a concern in the field of engineering, due to the scaling effect encountered in nanoscale devices. Therefore a necessary introduction follows:

In 1948, Casimir [1] made the remarkable prediction that two parallel, uncharged, perfect conductors in vacuum, would attract each other with a measurable force, if the separation between them is not too large. Even though this prediction is somewhat counter-intuitive from the point of classical electrodynamics, it in fact, was a culmination of research on forces between neutral objects (such as atoms or molecules) that started long ago with the work of van der Waals in 1873. In this short paper, an accurate history, as well as a derivation of the Casimir and van der Waals-type forces is not possible. We will just present a cursory look and so the reader is invited to study the excellent reviews and books devoted to these subjects [2–5].

Even though nearly 60 years have passed since Casimir's computation, there are still many things that we need to understand about the fundamentals of this force and its applications as well as its hazards in small electronic systems. There are two important facets of the Casimir force: first of all it is quantum mechanical and secondly, it is relativistic! Let us try to explain what we mean by these, but to make the discussion more concrete, let us recall that the attractive force between the plates reads

$$F = -\frac{\pi^2 \hbar c A}{240 d^4}, \qquad (1)$$

where $A$ is the surface area of the plates and $d$ is the separation distance between them. To simplify the analysis, we are assuming that the system under study is at a stationary point, namely, the plates are balanced by some other force and the system is in equilibrium.

As we can see $\hbar$, the reduced Planck's constant, and $c$, the speed of light enters into (1). The former showing its quantum mechanical- and the latter, its relativistic- origins. What is quite unusual about (1) is that the electric charge, which should determine the strength of coupling, does not appear (see ref. [10] for a discussion on this). In retrospect, tracing both quantum mechanical and relativistic origins of the Casimir force is in fact easy. In classical electrodynamics two neutral atoms or molecules can still interact with each other if they have permanent dipole moments. But up until the explanation of London [6], it was a great mystery as to why atoms/molecules without permanent dipole moments attracted each other with van der Waals forces. Classical Maxwell's theory abhors such an interaction in sharp contrast to the experiments done in gases. As a great success of quantum mechanics, London showed that even though certain atoms or molecules may not have permanent average dipole moments, the rules of quantum mechanics dictate that their instantaneous dipole moments fluctuate (above their average value, that is zero). This fluctuation leads to an attractive $1/r^7$ force between the atoms/molecules. London used non-relativistic quantum mechanics to derive his formula. Later it was realized that London-van der Waals theory does not correctly reproduce the forces between neutral molecules at large separations. In 1948 Casimir and Polder [7] took into consideration that it took virtual photons a finite time to carry the force between the molecules. Their computation in Quantum Electrodynamics led to a retarded van der Waals force that decreased as $1/r^8$ for large separations. Therefore, relativistic effects appear not as a result of fast moving atoms or molecules (or in the parallel plate case, the motion of plates) but simply as a result of large (above 10 nm) separation between the molecules (or plates).

The final formula obtained by Casimir and Polder was quite straightforward but the computation that led to it was quite complicated. Casimir wanted to remedy this and so sought a simpler derivation. This led him to consider the force between two parallel plates. In his computation he used the notion of zero-point fluctuations of the vacuum and how the existence of these parallel conductors affects the fluctuations. The basic idea behind this is the following: in any quantum field theory, free fields (such as electromagnetic fields) are considered as oscillators at each point of space. A quantum mechanical oscillator has a non-vanishing zero-point energy. Hence the vacuum (which is usually defined as the no-particle state) does not have a vanishing energy. In fact, it is just the opposite; it has a huge, if not a divergent, energy. Introducing conductive plates in vacuum changes this energy in such a way that there is a finite component that depends on the separation of the plates; hence the Casimir force appears as the derivative of this energy.

In 1956 Lifshitz [8] reconsidered the force between two neutral objects (they could be dielectric) from a totally different perspective. In this computation, instead of vacuum fluctuations, charge and current fluctuations (as a result of quantum mechanics) within the dielectrics were used. The Lifshitz theory is quite powerful and allows one to compute the forces more realistically as often encountered in experiments. For example if the space between the parallel plates is filled with a medium, then the force reads (see [4] for original references)

$$F = -\frac{\pi^2 \hbar c A}{240 d^4} \sqrt{\frac{\mu}{\varepsilon}} \left( \frac{2}{3} + \frac{1}{3\varepsilon\mu} \right) \qquad (2)$$

Showing the validity of the Casimir force experimentally had proved to be very difficult for the original parallel plate configuration [9]. There has been a revival of interest, in precise measurements of the Casimir force in areas ranging from cosmology to quantum electronics. It is quite possible that the recently found acceleration of the universe is caused by the vacuum energy.

Unwanted stiction between the components of micro/nano-electro-mechanical systems (MEMS/NEMS) can be caused by the Casimir force. And as discussed below, the Casimir force may influence the operation of large contact area molecular electronic devices in general, providing one unobvious reason for discrepancies in observed electrical characteristics; both from device to device and from architecture to architecture (i.e. as compared to single molecule devices). Most particularly switching junctions based on filamentary growth may be better understood by accounting for the Casimir force.

II. THEORETICAL APPROACH

We performed first order calculations for the Casimir force between a pair of planar metal electrodes and found it to be significant and sufficient to depress the electrodes and compress the molecular monolayer. Figure 1 depicts the magnitude of the Casimir pressure (force per unit area) as a function of plate separation, $d$, for plates separated by vacuum and a dielectric according to (1) and (2), respectively. As can be seen, the Casimir force is not substantially mitigated by the inclusion of a dielectric (molecular monolayer) between the plates.

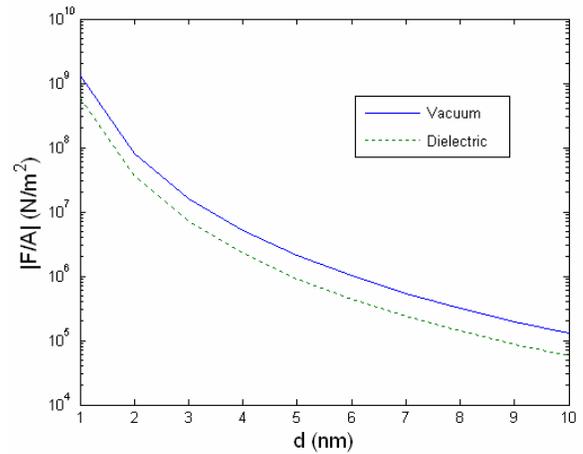

**Fig. 1** Magnitude of Casimir pressure as a function of distance between parallel plate conductors.

Reported values of Young's modulus for switching molecules range from 10 to 100 GPa making them as stiff as metals (~80 GPa). For a switching junction with a dimension of *1μm × 1μm* and a molecular monolayer thickness of *~3nm*, the Casimir force is *~16μN*. Our calculation shows that this force can cause more than 12% local compressive strain resulting in a localized monolayer compression of more than ~0.4nm. This ubiquitous force, found in all molecular electronic devices of this architecture, is compounded by the applied voltage bias between the metal electrodes. In filamentary switching junctions, this stimulates the generation of metal protrusions, leading to electrochemical reactions, thermal and/or electromigration, especially around local inhomogeneities. The presence of a high density of grains on freshly deposited metal surfaces significantly increases the Casimir attraction force because of the inherent nonlinear correlation with distance, causing even more device shorting given a rough electrode surface. Proper treatment of the Casimir force in designing molecular devices will shed light on some of the challenges including randomness in device characteristics and poor device yields.

## III. EXPERIMENTAL

Despite initial calculations of the Casimir force for a parallel plate architecture, the most successful experiments to date have shown the Casimir force via other architectures [11-14] (often employing a sphere and a plate). This stems from the difficulty of producing and maintaining a parallel separation between the plates throughout the course of the experiment. We propose a novel method for observing the Casimir force and showing its presence in molecular electronics based on the original parallel plate configuration.

Our preliminary experiments and data correspond to alkanethiol self-assembled monolayers (SAMs) fabricated on ultra-smooth Pt [15-16] bottom electrodes. From these experiments, it is noted that SAM orientation/disorder is detectable through reflection-absorption infrared spectroscopy (RAIR) measurements. Based on the selection rules, the RAIR spectroscopy of monolayers adsorbed on a metal surface shows that only the transition dipole perpendicular to the metal surface gives measurable absorption. Fig.2 shows an all-trans alkanethiol molecule on a metal surface. The tilt angle, $\alpha$, and twist angle, $\beta$, define the orientations of the chain with respect to the surface normal and the rotation of the plane containing the zigzag chain along the chain axis, respectively. It is noted that for the normal orientation, i.e. $\alpha = 0$, both the symmetric and asymmetric methylene vibrations ($v_s$ and $v_a$, respectively) are parallel to the metal plane, while vibrations of the methyl group have components perpendicular to the surface. Accordingly, the methylene groups ($CH_2$) in a perpendicular, all-trans alkyl chain will not exhibit absorption and the terminal methyl group does not show a difference in the projection along the surface normal. Once the chain is tilted from the surface normal (due to grains and roughness on the Pt surfaces), methylene groups of the alkyl chain are no longer parallel to the surface and thus will show a vibration absorption signature in its spectra. Therefore randomness in the orientation of the molecules can be detected via RAIR.

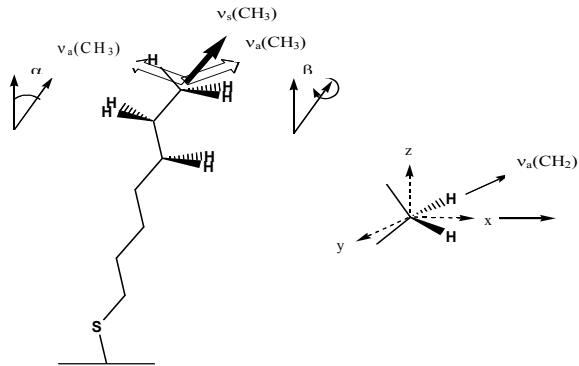

**Fig. 2**: Schematic diagram of chain geometry of an all-trans alkane chain. The tilt angle ($\alpha$) and the twist angle ($\beta$) are defined with respect to the surface normal and the rotation of the CCC plane relative to the surface normal and the chain tilt, respectively.

Fig.3 shows the RAIR spectra of monolayers of alkanethiols on a freshly deposited smooth Pt surface as well as a grainy Pt rough surface. The measured higher absorption in the RAIR spectrum for a randomly oriented monolayer adsorbed on the metal surface is evident in the figure. The band at 2965 cm$^{-1}$ is assigned to the $CH_3$ asymmetric in-plane CH stretching mode, $v_a(CH_3,ip)$, and the bands at 2937 and 2879 cm$^{-1}$ are assigned to the $CH_3$ symmetric stretching mode, $v_s(CH_3)$; this latter band is split due to the Fermi resonance interaction with the lower frequency asymmetric $CH_3$ deformation mode [17]. The bands at 2918 and 2850 cm$^{-1}$ are assigned to $v_a$ and $v_s$ $CH_2$ modes, respectively.

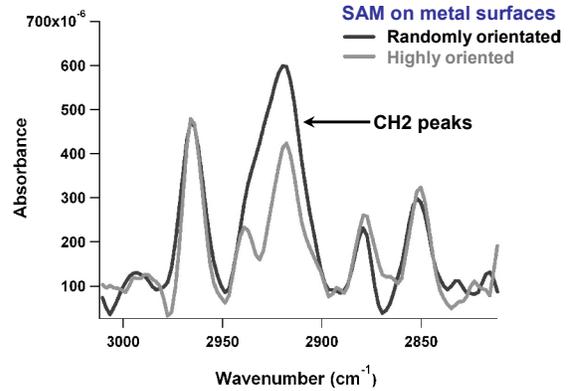

**Fig. 3** RAIR spectra of monolayers of alkanethiolates on an ultra-flat and a rough Pt surface.

In order to observe the Casimir force we are employing a new test structure comprised of self-assembled monolayers sandwiched between ultra-smooth (Ag, Au, Pt) electrodes. The top and bottom electrodes are fabricated according to a recently pioneered method for producing ultra-smooth Ag surfaces using a thin Ge nucleation layer [18, 19]. This new technique allows improvements in surface roughness, grain size and conductivity of the deposited metal electrode which will significantly enhance the measurement of the Casimir force in molecular electronics (see Fig. 4).

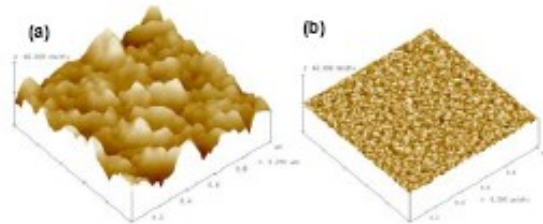

**Fig.4** AFM image of the (a) 15nm Ag and (b) 15nm Ag/1nm Ge. The Ag/Ge system significantly reduces the surface roughness and the grain size as compared to the Ag system (same height scale).

The bottom electrode is submersed in a dilute (5 mM) solution of 1,8-Octanedithiol (Sigma) for 36 hours under nitrogen ambient in order to produce the SAMs. The top electrode is fabricated using the same technique and gently placed on top of the bottom SAM-coated electrode to which it semi-covalently bonds. The new structures will be more robust than monothiol-based devices and the new electrodes confer smoother electrode surfaces.

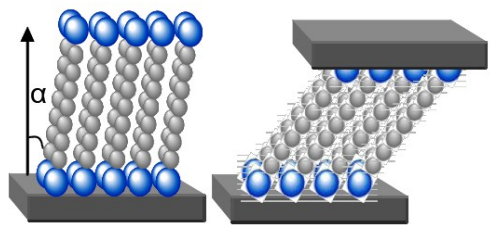

**Fig. 5** Schematic depicting increase of α (following top electrode application) resulting from the Casimir force.

RAIR spectra of the dithiol-based monolayers will be taken before and after the top electrodes are delineated. The Casimir force acting on the monolayer (after application of the top electrode) will serve to increase *α* and therefore the $CH_2$ absorbance peak relative to the reference sample. This change of *α* is depicted in Fig.5. Furthermore, some peak shifting are expected to occur as a result of compression along the carbon backbone of the molecule.

## IV. CONCLUSION

The Casimir force has been recognized as an inevitable and inherent force in developing nanoscale devices. Molecular electronics is one such area which can benefit immensely by accounting for its presence. Anomalies in experimental data, poor device yield due to electrical shorting, and conductance changes in molecular switching junctions can be better understood within the context of this often unaccounted force.

ACKNOWLEDGMENT

This ongoing work is partially supported by a NSF grant #0547679. B.T. was supported by the "Young Investigator Fellowship" of the Turkish Academy of Sciences (TUBA) and by the TÜBITAK Kariyer Grant No 104T17. The authors would like to thank Drs. Zhiyong Li, and Shun-Chi Chang of Hewlett-Packard Laboratories for their contributions in sample preparation and for helpful suggestions on the experiments